\documentclass[a4paper,twoside,notoc,11pt]{JHEP3}

\usepackage{epsfig,multicol}
\usepackage{delarray,amsmath,bbm}

\newcommand\fverb{\setbox\pippobox=\hbox\bgroup\verb}
\newcommand\fverbdo{\egroup\medskip\noindent%
                              \fbox{\unhbox\pippobox}\ }
\newcommand\fverbit{\egroup\item[\fbox{\unhbox\pippobox}]}
\newbox\pippobox

\newcommand{\beq} {\begin{equation}}
\newcommand{\eeq} {\end{equation}}
\newcommand{\beqa} {\begin{eqnarray}}
\newcommand{\eeqa} {\end{eqnarray}}

\newcommand{\ie}{{\it i.e.}}
\newcommand{\eg}{{\it e.g.}}

\newcommand{\as}{\alpha_s}

\newcommand{\order}[1]{${\cal O}\left(#1 \right)$}

\newcommand{\eq}[1]{(\ref{#1})}

\newcommand{\Pom}{\mathbb{P}}
\newcommand{\xP}{x_{\mathbb{P}}}
\newcommand{\x}{x_{B}}

\newcommand{\qu}{{\rm q}}
\newcommand{\qb}{{\rm\bar q}}
\newcommand{\qq}{\qu\qb\ }
\newcommand{\cc}{$c\bar c$ }

\newcommand{\etal}{{\it et al.}}
\newcommand{\PL}[3]{Phys.\ Lett.\ {{\bf#1}}, {#2} ({#3})}
\newcommand{\NP}[3]{Nucl.\ Phys.\ {{\bf#1}}, {#2} ({#3})}
\newcommand{\PR}[3]{Phys.\ Rev.\  {{\bf#1}}, {#2} ({#3})}
\newcommand{\PRL}[3]{Phys.\ Rev.\ Lett.\ {{\bf#1}}, {#2} ({#3})}
\newcommand{\ZP}[3]{Z. Phys.\ {{\bf#1}}, {#2} ({#3})}

\title{\center{Parton Propagation in a Gluon Field\thanks{Talk at the {\it Workshop on In-Medium Hadron Physics}, University of Giessen, November 2004.}}}
\author{Paul Hoyer\\
              Department of Physical Sciences and Helsinki Institute of
              Physics\\
              POB 64, FIN-00014 University of Helsinki, Finland \\
              E-mail: \email{paul.hoyer@helsinki.fi}}
\preprint{HIP-2005-06/TH\\  \hepph{0502207}
}
 
\abstract{The phenomenological success of PQCD is based on processes where the effects of the color field environment on parton propagation can be eliminated or is universal. In hard diffraction and quarkonium  production the PQCD subprocess is the same as in fully inclusive scattering, but the sensitivity to reinteractions is different. I discuss how this may be exploited to give new information on the dynamics of hard collisions.}

 
\begin{document}
 
\section{Introduction}\label{intro}
In this talk I address some issues concerning the medium effects on quark and gluon propagation. This is surely a simpler issue than in-medium effects on hadrons, which is the focus of this meeting. On the other hand, quarks and gluons are never free. If they are not bound in a hadron, partons find themselves in the quark and gluon condensate of the QCD vacuum. We know little of either environment -- yet have been able to make accurate predictions for hard processes by requiring that

\begin{itemize}

\item Partons are highly virtual and so do not propagate far enough to experience interactions with the environment. This is the case in hard QCD subprocesses.

\item When partons are nearly on-shell, they move so fast through hadron targets that only elastic Coulomb scattering can occur. Such interactions combine with the target wave function into the measured parton distributions.

\end{itemize}

The above conditions are more precisely formulated in the QCD factorization theorems \cite{factorization}, and have allowed us to establish QCD as the theory of the strong interaction. Given this success, questions which go beyond the standard factorization framework arise:

\begin{itemize}

\item How can we learn about the color field environment of hard collisions?

\item How does the rescattering of a struck parton on spectators influence the measured parton distribution?

\item How does the QCD vacuum affect the propagation of quarks and gluons, causing confinement? 

\end{itemize}

In the following I address some aspects of the above issues. {\em Diffractive DIS} requires a parton rescattering to turn the target spectator system into a color singlet. {\em Quarkonium production} is sensitive to the color field environment since some states ($J/\psi,\chi_1$) couple to a minimum of three gauge bosons, while others ($\eta_c,\chi_2$) make do with two. Studies of Green functions in a fluctuating {\em background color field} may provide insight into the effects of the QCD vacuum on the propagation of quarks and gluons\footnote{I refer to \cite{Hoyer:2004iw} for this part of my talk, which cannot be included here due to the space limit of the proceedings.}.

\begin{figure}[tb]
\centerline{\epsfig{file=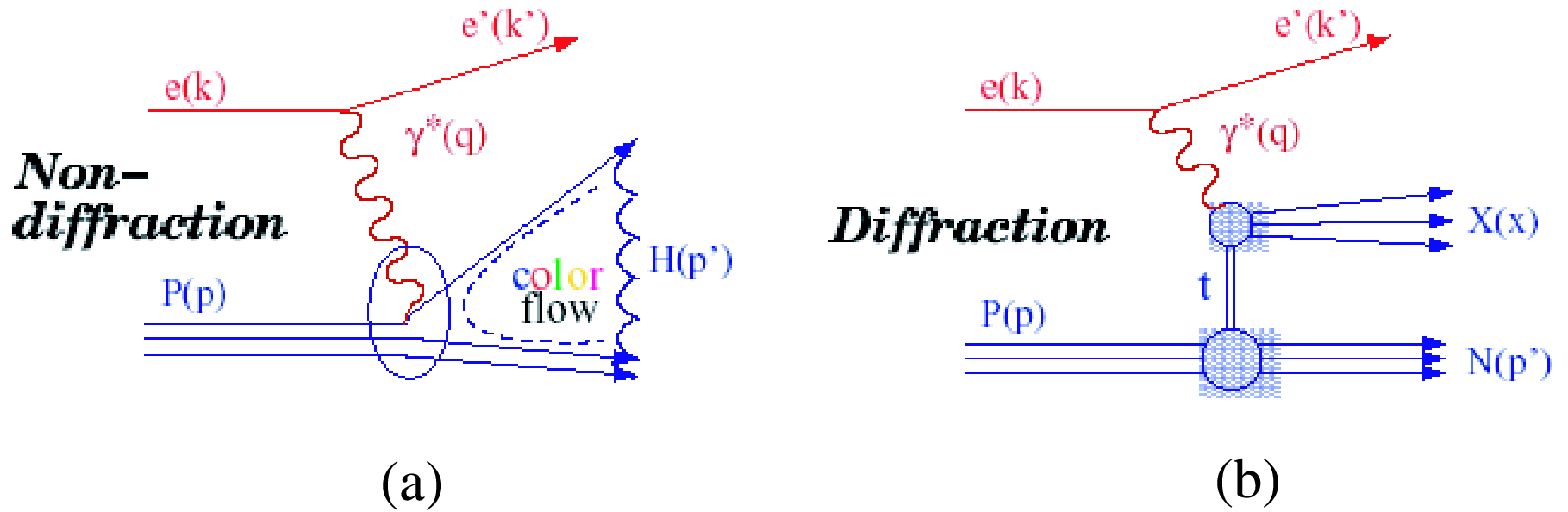,width=.9\columnwidth}}
\caption{Non-diffractive (a) and diffractive (b) Deep Inelastic Scattering.
\label{fig1}}
\end{figure}

\section{Diffractive DIS}

In Deep Inelastic Scattering (DIS) of leptons on nucleons the virtual photon strikes a quark (or gluon) out of the target. The separation of the struck parton from the target spectator system is assumed to generate a color string, the breaking of which produces hadrons at all rapidities in the final state (Fig.~1a). This picture is in qualitative agreement with about 90\% of DIS events. However, in the remaining 10\% there is a large rapidity gap in the final state \cite{Abramowicz:2004dt}, with the target nucleon often surviving intact (Fig.~1b). This Diffractive DIS (DDIS) process is apparently of `leading twist', \ie, the ratio $\sigma(DDIS)/\sigma(DIS) \simeq 0.1$ is not suppressed by a power of $Q^2$.

In an early model of DDIS due to Ingelman and Schlein (IS) \cite{IS} (which preceded the data) the photon scatters from a color singlet component of the target proton wave function, the `Pomeron', which carries a small fraction $\xP$ of the proton momentum. This picture implies that the fraction of diffractive events should be about 10\% in all hard processes. However, the measured fraction of diffractive events in hadron induced processes is only $\sim 1\%$ \cite{Abramowicz:2004dt}. The energy dependence of the lepton scattering data \cite{Breitweg:1998gc} is moreover incompatible with a Regge picture:
\beq \label{disratio}
\frac{d\sigma_{diff}^{\gamma^*p}/dM_X}{\sigma_{tot}^{\gamma^*p}} \propto
\frac{(W^2)^{2\bar\alpha_\Pom-2}}{(W^2)^{\alpha_\Pom-1}}  \propto \left\{ 
\begin{array}{ll}W^{0.19} & {\rm Regge} \\ W^{0.00 \pm 0.03} & {\rm ZEUS} \end{array} \right.
\eeq
where $W$ is the total mass of the hadronic system, $M_X$ is the mass of the diffractive system and $\bar\alpha_{\Pom}$ is the average value of the Pomeron trajectory for the relevant range of momentum transfer. From a theoretical point of view, the IS picture of the Pomeron as a constituent of the initial target wave function is qualitatively different from our usual understanding of diffraction as the ``shadow'' of inelastic channels -- which requires the amplitude to have an absorptive part.

A large rapidity gap is possible only when the hadronic systems on either side of the gap are color singlets. The DDIS dynamics must therefore involve a color neutralization of the target remainder, which should happen before hadronization (and thus color string formation) has had time to begin. QCD in fact offers a mechanism for this: soft Coulomb rescattering of the struck parton on its way out of the target \cite{BHMPS,Brodsky:2004hi}. Such rescattering can occur instantaneously and within the `Ioffe' coherence length of the virtual photon ($L_I \simeq 1/2mx_B$) where it cannot be resolved from the hard vertex.

Parton rescattering is an integral part of DIS dynamics and is described by the path ordered exponential or `Wilson line'
\beq \label{POE}
W[x^-;0] = {\rm P}\exp\left[ig\int_0^{x^-} dw^- A^+(w^-) \right] 
\eeq
in the target matrix element that gives the usual quark distribution,
\beq
f_{\qu/N}(\x,Q^2) = \frac{1}{8\pi} \int dx^- \exp(-i \x p^+ x^-/2)
 \langle N(p)|\bar\psi(x^-) \gamma^+\, W[x^-;0] \, \psi(0)|N(p)\rangle\  \label{PDFdef} 
\eeq
The measured parton distributions thus reflect both the initial presence of partons in the target wave function and the rescattering of the struck parton in the color field of the target spectators. The rescattering gives rise to dynamical phases (arising from on-shell intermediate states) and interference effects. These manifest themselves in the observed nuclear shadowing \cite{BHMPS}, diffraction \cite{Brodsky:2004hi} and single spin asymmetries \cite{BHS}.

A perturbative illustration of the rescattering dynamics generated by the Wilson line \eq{POE} is shown in Fig. 2a. The target quark emits a gluon which carries a small momentum fraction $\sim ``\xP$''. The gluon, in turn, fluctuates into a $\qq$ pair. The rapidity separation of the quark pair from the target, $\sim \log1/\xP$, will correspond to the rapidity gap in a diffractive event. The $\qq$ pair has a large transverse size $\sim 1$ fm since the gluon has low virtuality.

\begin{figure}[htb]
\centerline{\epsfig{file=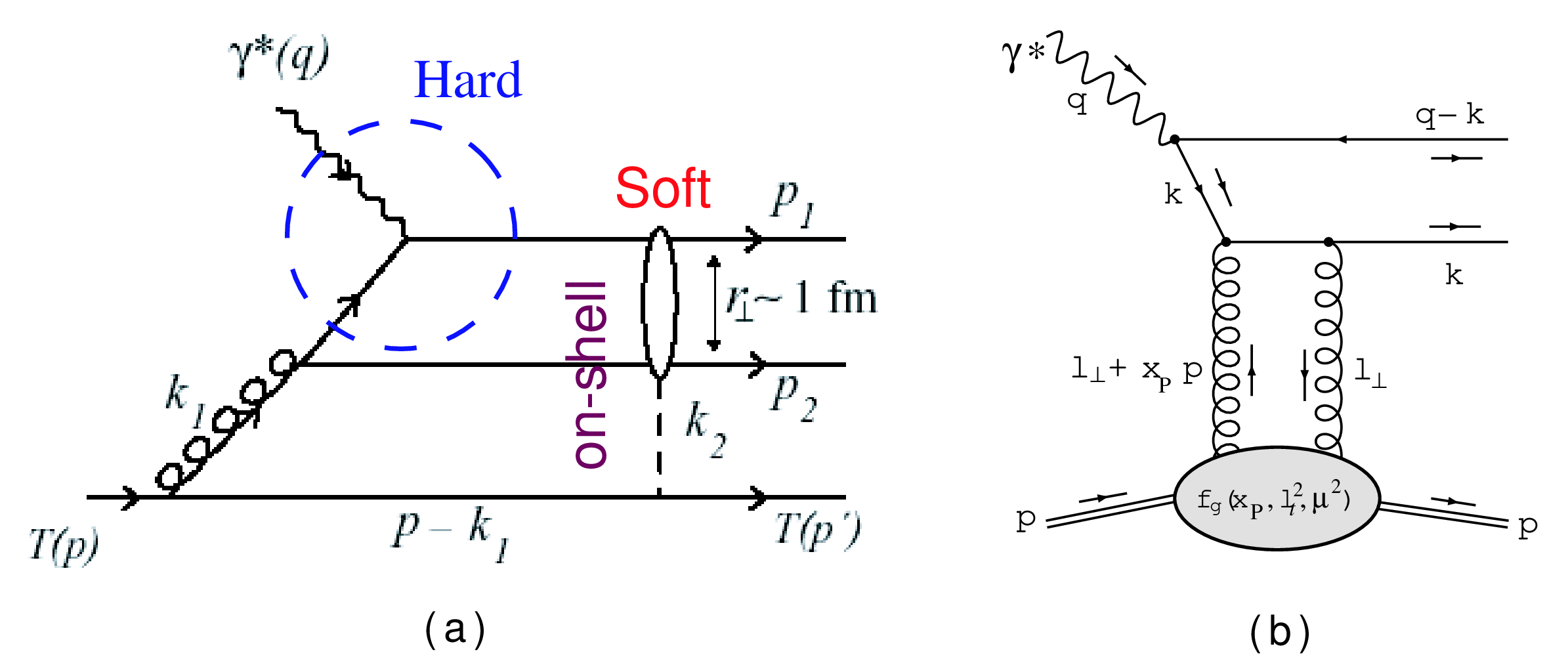,width=.9\columnwidth}}
\caption{Perturbative models of rescattering dynamics. In (a), the struck quark $p_1$ forms a color dipole together with the sea quark $p_2$, which is separated by a large rapidity gap from the target quark $p'$. The soft rescattering $k_2$ proceeds instantaneously via Coulomb gluon exchange. In diagram (b), taken from \protect{\cite{Martin:2004xw}}, both gluons have erroneously been assumed to be transverse and the lower vertex to be a Generalized Parton Distribution. \label{fig2}} 
\end{figure}

Next the virtual photon strikes one of the quarks, giving it a huge momentum. The struck quark ($p_1$) Coulomb scatters on the target ($A^+$ exchange indicated by the dashed line $k_2$). The companion quark ($p_2$) also has large longitudinal momentum with respect to the target (due to the small $\xP$) and may similarly Coulomb scatter. The relative contribution of the rescattering from each quark is gauge dependent: In Feynman gauge only the scattering of $p_1$ is relevant, whereas in light-cone gauge ($A^+=0$) the full contribution comes from the scattering of $p_2$ (via a singular term in the light-cone gauge propagator). In Fig. 2a the gauge invariant sum of the rescatterings of both quarks is indicated by a blob. 

The Coulomb exchange $k_2$ is soft since the quark pair is in a color octet state (this can be a `monopole' exchange in the terminology of the next section). Such exchanges are not suppressed, there can be many of them and they cannot be reliably evaluated using perturbation theory. All exchanges which occur within the Ioffe distance $L_I \simeq 1/2mx_B$ of the hard vertex are coherent with the virtual photon interaction and thus affect the DIS cross section. 

Due to the large transverse size ($\sim 1$ fm) of the quark pair ($p_1,p_2$) the soft gluon exchange $k_2$ can couple also to its color dipole moment. As in our discussion below for quarkonium, the quark pair can then be transformed into a color singlet. If there are no further exchanges the pair separates from the target without the formation of a color string: A rapidity gap is formed between $p_2$ and $p'$ in Fig. 2a and the event is classified as diffractive (DDIS).

In Fig. 2a the hard virtual photon interaction is drawn at lowest order in $\as$. Higher orders  also contribute via vertex corrections and hard gluon emission. These do not affect the soft rescattering $k_2$, which due to its low transverse momentum does not resolve the compact substructure of the hard vertex. The transverse distance between the struck quark and the hard emitted gluons does not have time to grow inside the target, due to Lorentz time dilation. This ensures the factorization between the hard subprocess and the target parton distributions (which include rescattering effects).

The higher order corrections also do not affect the presence of a rapidity gap in diffractive DIS. Straightforward estimates show \cite{Brodsky:2004hi} that the hard gluons are emitted with rapidities between those of $p_1$ and $p_2$, \ie, the hard radiation does not enter the rapidity gap. This means that the DGLAP evolution of diffractive and inclusive parton distributions is the same, as required by the DDIS factorization theorem \cite{diff-factorization}. 

The two-gluon exchange mechanism of Fig. 2a differs in important respects from models \cite{Martin:2004xw} based on Fig. 2b, where the lower vertex is given by the (generalized) parton distribution (GPD) of the target. In a GPD both gluons are transversely polarized and couple to a transversally compact object (the difference between the transverse coordinates of the quark fields in Eq. \eq{PDFdef} is of \order{1/Q}). The GPD diagram is correct for {\it exclusive} diffraction, such as deeply virtual meson production \cite{Abramowicz:2004dt}, where the quark pair is compact. The exclusive contributions to the DIS and DDIS cross sections are, however, suppressed by powers of $1/Q^2$ since both gluons are hard. The logarithmic scaling violations of the parton distributions given by the GPD in Fig. 2b (which enters squared in the cross section) would also be different from the standard DGLAP behavior mandated by the above discussion and the DDIS factorization theorem \cite{diff-factorization}.

\subsection*{Hadron-Induced Diffractive Processes}

As I already remarked above, the fraction of events with a rapidity gap is much lower ($\sim 1\%$) in hadron induced processes (such as $pp$) as compared to DDIS. This non-universality means that diffraction cannot occur simply through scattering on a ``Pomeron'' component in the target wave function, as postulated in the early IS model. It is furthermore recognized that the universality of diffractive parton distributions in QCD does not extend to diffractive processes induced by hadron beams  \cite{diff-factorization}.

\EPSFIGURE[h]{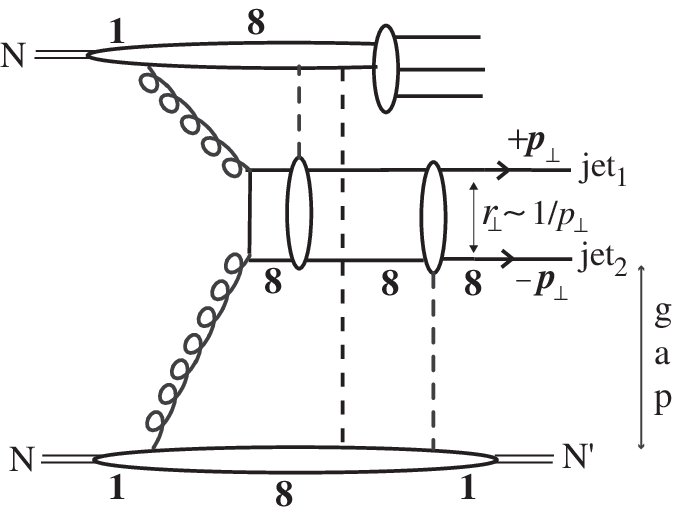,width=0.4\columnwidth}{Illustration of diffraction through rescattering in
$NN \to 2\ jets + X$ in analogy with, and using  
the same notation as, the DIS case in Fig.~\protect\ref{fig2}. 
The compact $\qq$ pair which forms the jets is assumed to be in a color octet ({\bf 8}) configuration. This pair rescatters coherently  and thus retains its color.\label{fig3}}


An essential difference between virtual photon and hard hadron diffraction is that the hadron spectator system is colored. There are then many more possibilities for Coulomb rescattering to occur, as illustrated in Fig. 3. Consequently the probability for a rapidity gap to arise is process-dependent. Systematic comparisons of diffractive parton distributions measured in different processes can give us information on the color field environment of hard processes, which is inaccessible in fully inclusive processes.

As in DDIS, the hard radiative corrections to the partonic subprocess should affect neither the rescattering probability nor the presence of a rapidity gap. Hence the scaling violations of diffractive parton distributions will be governed by the usual DGLAP equations also in hadron induced diffraction.

The fact that the measured DDIS/DIS cross section ratio \eq{disratio} is approximately independent of $x_B$ implies that the constraint on rescattering in DDIS (that the target system is left in a color singlet state) does not significantly affect the overall momentum transfer from the target. In PQCD diagrams the longitudinal momentum fraction carried by the Colomb rescattering gluons ($k_2$ in Fig. 2a) is generally similar to that of the primary gluon ($k_1$). However, PQCD cannot be trusted to quantitatively describe the soft rescattering exchanges. It will be interesting to measure the $x_B$-dependence of diffractive parton distributions also in $pp$ collisions, and to compare their shape to those in inclusive hard scattering.

\subsection*{Similarity to the SCI model}

The QCD description of hard diffractive scattering that I have discussed here is qualitatively similar to the ``Soft Color Interaction'' (SCI) model developed by Ingelman and collaborators \cite{SCI}. In the SCI Monte Carlo the hard subprocess is the same as in usual inclusive processes. Before hadronization begins, soft gluon exchanges are postulated to redistribute color between the spectator system(s) and the products of the hard subprocess. These exchanges carry no momentum and are qualitatively similar to the Coulomb rescatterings which we now recognize to be a property of QCD. The SCI model is able to describe the measured distribution of rapidity gaps with a single new parameter, the probability for a soft exchange to occur between two partons. In particular, the observed relative suppression of diffractive events in hadron induced processes as compared to DDIS is reproduced in SCI using the same rescattering probability in both processes.

Some aspects of the SCI model differ from QCD expectations. In particular, the color coherence effects in scattering from compact subsystems are not taken into account. Since Monte Carlo simulations are probabilistic they do not incorporate quantum mechanical interference, but it should be possible to model those effects in a qualitative sense.

We appear to be at a threshold of very promising developments in our understanding of the dynamics of hard diffraction. Future data should be able to test whether the hard subprocesses really are the same in diffractive and inclusive processes, including their  radiative corrections which imply universal scaling violations in diffractive and inclusive parton distributions. Systematic comparisons of diffractive parton distributions in different processes (DDIS, $pp$, events with one or several rapidity gaps...) should allow us to learn about the color environment in hard processes, which we have so successfully avoided to address in fully inclusive reactions.

\section{Quarkonium production}

The QCD factorization theorems \cite{factorization} allow us to predict sufficiently inclusive quantities, \eg, total charm quark production or $D$-meson production integrated over the intrinsic transverse momentum in the jet. The effects of soft interactions with the color field environment are small or universal in such quantities. On the other hand, standard factorization does not apply to quarkonium production, which is a small and `fragile' part of the total heavy quark production cross section. The fact that quarkonium production is thus a sensitive `thermometer' was recognized long ago and is extensively utilized in the study of heavy ion collisions \cite{Bedjidian:2003gd}. Many aspects of the quarkonium production mechanism remain mysterious also in collisions with hadron projectiles \cite{Hoyer:1998dr}.

\begin{figure}[htb]
\centerline{\hspace{1cm}\epsfig{file=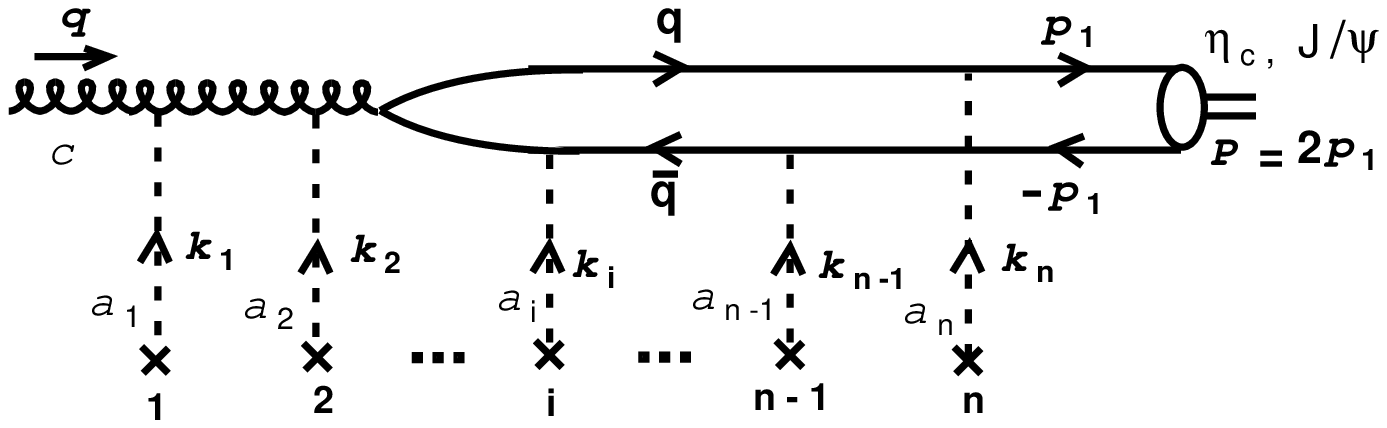,width=.9\columnwidth}}
\caption{Production of charmonium states through high energy gluon splitting $g \to c\bar c$ and $n$-fold rescattering from static color sources. From Ref. \cite{Hoyer:1997yf}.  \label{fig4}} 
\end{figure}

Quarkonium production is particularly sensitive to rescattering effects since the heavy quark pair is produced with specific quantum numbers. This is illustrated in Fig. 4 where a high energy gluon splits into a \cc pair, which materializes as $\eta_c\ (J^{PC}=0^{-+})$ or $J/\psi\ (1^{--})$ charmonium \cite{Hoyer:1997yf}. At least one rescattering is required to turn the incoming gluon into color singlet charmonium, and this has to occur off the \cc pair rather than off the projectile gluon (since a gluon always remains a color octet).

In QCD processes like those of Fig.~4 one needs to distinguish between color {\it monopole} and color {\it dipole} scattering. A single gluon scatters as a monopole, whereas the \cc pair in Fig.~4 also has a dipole contribution, characterized by an extra factor of transverse size $r_\perp \sim 1/m_c$ in the amplitude. For compact objects like a heavy quark pair this suppresses the cross section and makes dipole scattering hard.

It turns out \cite{Hoyer:1997yf} that the final gluon ($k_n$ in Fig.~4) is always hard, corresponding to dipole exchange. This is intuitively easy to understand, as that gluon turns the \cc pair from color octet to singlet and must thus be able to resolve the quarks. In the case of $\eta_c$ production all earlier exchanges ($k_1\ldots k_{n-1}$) are (dominantly) soft monopole exchanges, which leave the \cc in a color octet state. For the $J/\psi$ we know that the \cc pair must couple to a minimum of three gluons. The final one ($k_n$) should be a dipole exchange for the reason already mentioned, but it is not {\it a priori} clear whether it suffices that the third (and more) gluons are soft, monopole exchanges. The calculation showed that precisely one more gluon must be hard, and it can be any of the other exchanges $k_i\ldots k_{n-1}$ coupling to the \cc pair. All the remaining exchanges can be soft. Thus the requirement of charge conjugation is that a minimum of three gluons must couple to the \cc  {\it dipole} to produce a $J/\psi$.

In the above analysis the momentum of the incoming gluon was assumed to be asymptotically large, hence the transverse size $r_\perp \sim 1/m_c$ of the \cc state was frozen during rescattering. This is analogous to the ``Color Singlet Model'' (CSM) \cite{csm}, where a third gluon is {\it emitted} at an early phase of the production process. Conceivably, the third gluon could be emitted much later in the $J/\psi$ formation process, when the \cc pair has expanded to the size $r_\perp \sim 1/\alpha_sm_c$ of the bound state. The dipole suppression factor would then be less significant, giving an enhanced cross section. The ``Color Octet Model'' (COM) \cite{com} and the more systematic NRQCD expansion \cite{nrqcd} use this approach to explain why the hadronic production cross section of the $J/\psi$ is more than an order of magnitude larger than the CSM prediction.

There are, however, features of the quarkonium production data which are not expected in the NRQCD approach. For example, $J/\psi$  {\it photoproduction} is very well described by the CSM (at NLO in $\as$) \cite{csmphoto}. In NRQCD the soft emission should be qualitatively independent of the hard \cc production process, hence one expects similar enhancements for hadroproduction and photoproduction. The {\it nuclear target dependence} scales in the Feynman momentum fraction $x_F$ of the $J/\psi$ rather than in the target parton's $x_2$, as would follow from QCD factorization \cite{Hoyer:1990us}. These and other features of the data on quarkonium production for a variety of beams and targets prompted us to propose \cite{Hoyer:1998dr} the rescattering scenario shown if Fig.~5.

\begin{figure}[hbt]
\centerline{\epsfig{file=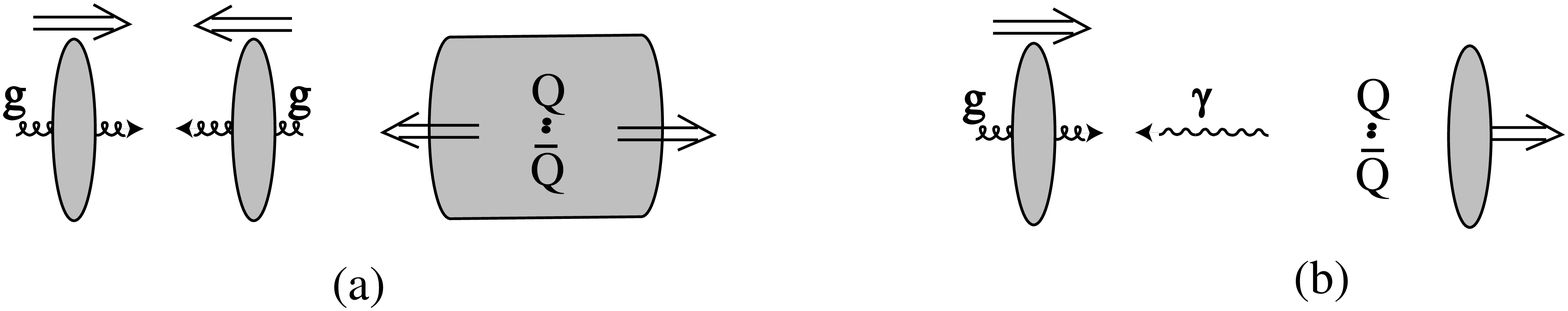,width=.9\columnwidth}}
\caption{Creation of heavy quarks in hadroproduction (a) and photoproduction (b). The color radiation fields carried by the gluons in (a) may interact, forming a color field at the same rapidity as the heavy quarks. Quark pairs formed in the photon hemisphere in (b) have no  comoving color fields.
\label{fig5}}
\end{figure}

Gluons are accompanied by a color field which is radiated in hard processes, such as in heavy quark production. In hadroproduction (Fig.~5a, $gg \to Q\bar Q$) the radiation fields of the two gluons may interact to create a color field around the heavy quark pair. In photoproduction (Fig.~5b, $\gamma g \to Q\bar Q$) the photon carries no color field, consequently a heavy quark pair created in the photon fragmentation region sees no color field. This could explain why the third gluon must be emitted in photoproduction (as in the CSM), whereas interactions with the surrounding field in hadroproduction would allow to satisfy the requirements of charge conjugation without gluon emission. This scenario can qualitatively explain several aspects of the data, but quantitative predictions remain model-dependent.

\section*{Acknowledgments}

This talk was based on work done in collaboration with Stan Brodsky, Rikard Enberg, Gunnar Ingelman and St\'ephane Peign\'e. I am also grateful to the organisers of this meeting for their kind invitation. My work is partially supported by the Academy of Finland through grant 102046.


\begin{thebibliography}{99}

\bibitem{factorization}
J.~C.~Collins and D.~E.~Soper,
Nucl.\ Phys.\ B {\bf 194}, 445 (1982);
%
J.~C.~Collins, D.~E.~Soper and G.~Sterman,
Nucl.\ Phys.\ B {\bf 261}, 104 (1985);
%
Nucl.\ Phys.\ B {\bf 308}, 833 (1988)
and
Phys.\ Lett.\ B {\bf 438}, 184 (1998)
[arXiv:hep-ph/9806234];
%
G.~T.~Bodwin,
Phys.\ Rev.\ D {\bf 31}, 2616 (1985);
[Erratum-ibid.\ D {\bf 34}, 3932 (1986)].

\bibitem{Hoyer:2004iw}
P.~Hoyer and S.~Peigne,
JHEP {\bf 0412} (2004) 051
[arXiv:hep-ph/0410235].

\bibitem{Abramowicz:2004dt}
H.~Abramowicz,
arXiv:hep-ex/0410002.

\bibitem{IS}
G.~Ingelman and P.~E.~Schlein,
Phys.~Lett.~{\bf 152B}, 256 (1985).

\bibitem{Breitweg:1998gc}
J.~Breitweg {\it et al.}  [ZEUS Collaboration],
Eur.\ Phys.\ J.\ C {\bf 6}, 43 (1999)
[arXiv:hep-ex/9807010]; 
%
  [ZEUS Collaboration],
arXiv:hep-ex/0501060.


\bibitem{BHMPS}
S.~J.~Brodsky, P.~Hoyer, N.~Marchal, S.~Peign\'e and F.~Sannino,
Phys.\ Rev.\ D {\bf 65}, 114025 (2002)
[arXiv:hep-ph/0104291].

\bibitem{Brodsky:2004hi}
S.~J.~Brodsky, R.~Enberg, P.~Hoyer and G.~Ingelman,
arXiv:hep-ph/0409119.

\bibitem{BHS}
S.~J.~Brodsky, D.~S.~Hwang and I.~Schmidt,
Phys.\ Lett.\ B {\bf 530}, 99 (2002)
[arXiv:hep-ph/0201296].

\bibitem{Martin:2004xw}
A.~D.~Martin, M.~G.~Ryskin and G.~Watt,
Eur.\ Phys.\ J.\ C {\bf 37} (2004) 285
[arXiv:hep-ph/0406224].

\bibitem{diff-factorization}
J.~C.~Collins,
Phys.~Rev.~D {\bf 57}, 3051 (1998);
Erratum:\ \emph{ ibid.}\ D {\bf 61}, 019902 (1998)
[arXiv:hep-ph/9709499].

\bibitem{SCI}
A.~Edin, G.~Ingelman and J.~Rathsman,
Phys.~Lett.~B {\bf 366}, 371 (1996)
[arXiv:hep-ph/9508386];
%
Z.~Phys.~C {\bf 75}, 57 (1997)
[arXiv:hep-ph/9605281].

\bibitem{Bedjidian:2003gd}
M.~Bedjidian {\it et al.},
arXiv:hep-ph/0311048;
M.~J.~Leitch  [FNAL E866/NuSea and PHENIX Collaborations],
Eur.\ Phys.\ J.\ A {\bf 19} (2004) SUPPL1129.

\bibitem{Hoyer:1998dr}
P.~Hoyer,
Nucl.\ Phys.\ Proc.\ Suppl.\  {\bf 75B} (1999) 153
[arXiv:hep-ph/9809362]; 
%
P.~Hoyer and S.~Peigne,
Phys.\ Rev.\ D {\bf 59} (1999) 034011
[arXiv:hep-ph/9806424]; 
%
N.~Marchal, S.~Peigne and P.~Hoyer,
Phys.\ Rev.\ D {\bf 62} (2000) 114001
[arXiv:hep-ph/0004234].

\bibitem{Hoyer:1997yf}
P.~Hoyer and S.~Peigne,
Phys.\ Rev.\ D {\bf 57} (1998) 1864
[arXiv:hep-ph/9706486].

\bibitem{csm} J. H. K\"uhn, \PL{89B}{385}{1980}; C. H. Chang,
\NP{B172}{425}{1980}; E. L. Berger and D. Jones, \PR{D23}{1521}{1981}; R.
Baier and R. R\"uckl, \PL{102B}{364}{1981} and
\ZP{C19}{251}{1983}.

\bibitem{com} E. Braaten and S. Fleming, \PRL{74}{3327}{1995},
hep-ph/9411365; M. Cacciari, M. Greco, M. L. Mangano and
A. Petrelli, \PL{B356}{553}{1995}, hep-ph/9505379; P. Cho and A. K.
Leibovich, \PR{D53}{150}{1996}, hep-ph/9505329 and \PR{D53}{6203}{1996},
hep-ph/9511315.

\bibitem{nrqcd} G. T. Bodwin, E. Braaten and G. P. Lepage,
\PR{D51}{1125}{1995}, Erratum {\em ibid.,} {\bf D55}, 5853 (1997),
hep-ph/9407339.

\bibitem{csmphoto} M. Kr\"amer, J. Zunft, J. Steegborn and P. M. Zerwas,
\PL{B348}{657}{1995}; M. Kr\"amer, \NP{B459}{3}{1996}; 
%
S. Aid \etal\ (H1 Collaboration), \NP{B472}{3}{1996} 
[arXiv:hep-ex/9603005]; 
%
J. Breitweg \etal\ (ZEUS Collaboration) 
\ZP{C76}{599}{1997} [arXiv:hep-ex/9708010].

\bibitem{Hoyer:1990us}
P.~Hoyer, M.~V\"anttinen and U.~Sukhatme,
Phys.\ Lett.\ B {\bf 246} (1990) 217.

\end{thebibliography}
\end{document}